\documentclass{aa}
\usepackage{graphicx}

\begin{document}

\title{Detecting cold H$_2$ globules in the outer Galactic disc by
microlensing towards the \object{Maffei~1} elliptical}

\author{R. Fux}

\institute{Geneva Observatory, Ch. des Maillettes 51,
           CH-1290 Sauverny, Switzerland\\
           \email{Roger.Fux@obs.unige.ch}}

\date{Received xx / Accepted 20 September 2004}

\abstract{A serious candidate for dark matter in spiral galaxies are
cold molecular hydrogen globules with a condensed central core and a
disc-like space distribution probably similar to that of neutral
hydrogen. This paper shows that the H$_2$ cores are sufficiently
compact and massive to be detected by microlensing in the outer
Galactic disc and that the Maffei~1 elliptical galaxy, at a distance
of 3~Mpc and Galactic latitude $b=-0.6^{\circ}$, offers an ideal
target for such an experiment. The microlensing optical depth of H$_2$
cores along the line of sight to this galaxy is estimated to
$\tau\sim 0.7\cdot 10^{-6}$ if most of the dark mass in the Milky Way
resides in such cores, and the typical event timescale to
$\la 1$~day. Detection rates are computed both in the classical and
pixel lensing approaches in the $I$- and $K$-bands, and for a
representative selection of existing observing facilities. In the more
efficient pixel lensing case, two 10-hour observing runs, separated
in time by at least several days, should yield of order 10~positive
detections at the $5\sigma$ level using ground-based 8m-class
telescopes in the $K$-band or the Hubble Space Telescope~ACS camera in
the $I$-band, and the corresponding fraction of events with timescale
measurable to an accuracy better than $50\%$ amounts to about $9\%$
and $4\%$ respectively for these observing alternatives.
}
\titlerunning{Detecting cold H$_2$ globules in the outer Galactic disc
by microlensing towards Maffei~1}
\maketitle

\section{Introduction}
%
The nature of dark matter (DM) in the universe is one of the main
persistent problem in modern astrophysics. In spiral galaxies, dark
matter is highlighted by flat HI rotation curves extending far beyond
the optical disc, where one would expect a Keplerian fall-off
(e.g. Rogstad \& Shostak~\cite{RS}). Many DM candidates have been
proposed in the literature, including massive compact objects,
neutrinos and other more exotic particles. Compact objects have been
and are still searched for in the local group using classical
or pixel microlensing of stellar sources located in the
Galactic bulge, the Magellanic Clouds and Andromeda (e.g. the OGLE,
MACHO, EROS, MOA, AGAPEROS, MEGA, POINT-AGAPE, WeCAPP
collaborations\footnote{See respectively Wozniak et al.~\cite{OGLE},
Alcock et al.~\cite{MACHO}, Afonso et al.~\cite{EROS}, Bond et
al.~\cite{MOA}, Melchior et al.~\cite{AGAP}, de Jong et
al.~\cite{MEGA}, Paulin-Henriksson et al.~\cite{POINT}, and Riffeser
et al.~\cite{WECAPP} for selected references.}).
All these experiments are designed to find dark objects in the inner
Galaxy or in the more or less spherical halos around the Milky Way and
Andromeda galaxies.
\par An attractive possible constituent of galactic DM is H$_2$ in the
form of cold ``globules'', made of a condensed solid or liquid central
core surrounded by an extended atmosphere (Pfenniger~\cite{P};
Pfenniger \& Combes~\cite{PC}; see also White~\cite{RSW}). These H$_2$
cores would have radii of up to $2\cdot 10^4$~km and masses up to the
Earth mass. A number of constraints, like the \hbox{HI--DM} relation
(Bosma~\cite{BA1}), the disc-halo conspiracy, the maximum disc
property implying hollow spherical halos, and the large angular
momentum of accreted high velocity clouds, do support a space
distribution of H$_2$ globules similar to the HI distribution, i.e. as
an outer disc around spiral galaxies. Evidence for large amounts of
molecular gas in such regions is also provided by the observed star
formation in the extreme outskirts of the HI disc in M31 (Cuillandre
et al.~\cite{CLA}). Moreover, \hbox{$N$-body} simulations show that
massive collisionless DM discs are prone to bending instabilities
leading to long lasting warps (Revaz \& Pfenniger 2004), consistent
with the observed high frequency of these structures.
\par This paper shows that if H$_2$ cores contain most of the
Milky~Way's DM and have a disc-like distribution, they should be
quite easily detectable through a short microlensing experiment
targeting the low Galactic latitude Maffei~1 galaxy.
\par The structure of the paper is as follow. Section~\ref{maf}
summarises the main observational characteristics of Maffei~1.
Section~\ref{basic} briefly outlines the microlensing aspects useful
to our investigation and evaluates the basic properties resulting from
the lensing of extragalactic sources by H$_2$ cores. Section~\ref{lum}
introduces the concept of fluctuation magnitude and attempts to derive
a realistic stellar luminosity function for Maffei~1.
Section~\ref{rate} then computes the expected event detection rates
for a representative set of available observational facilities, both
in the classical and pixel lensing workframes. Finally,
Sect.~\ref{constr} argues that a massive Galactic H$_2$-globule disc
is not excluded by current microlensing constraints, and
Sect.~\ref{concl} concludes the paper. The problem of contamination by
variable stars is beyond the scope of this paper and will not be
treated, and the case of more diffuse H$_2$ clouds is analysed in
Rafikov \& Draine~(\cite{RD}). Much of the notations and theory
elements are based on Gould~(\cite{G96}; hereafter G96).
\begin{figure}[t!]
\centerline{\includegraphics[width=7cm]{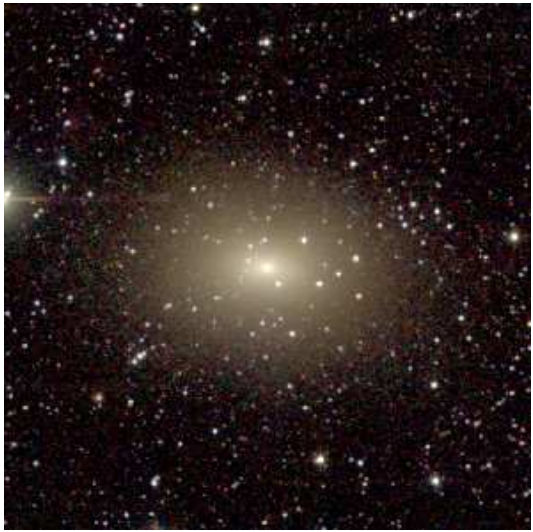}}
\caption{\small
$JHK$ composite image of Maffei~1 from the 2MASS Large Galaxy
Atlas (Jarrett et al.~\cite{JCC}). The field of view is
$11.9'\!\times\! 11.9'$.}
\label{obs}
\end{figure}

\section{The Maffei~1 galaxy}
\label{maf}

\begin{figure}[t!]
\centerline{\includegraphics[width=7.8cm]{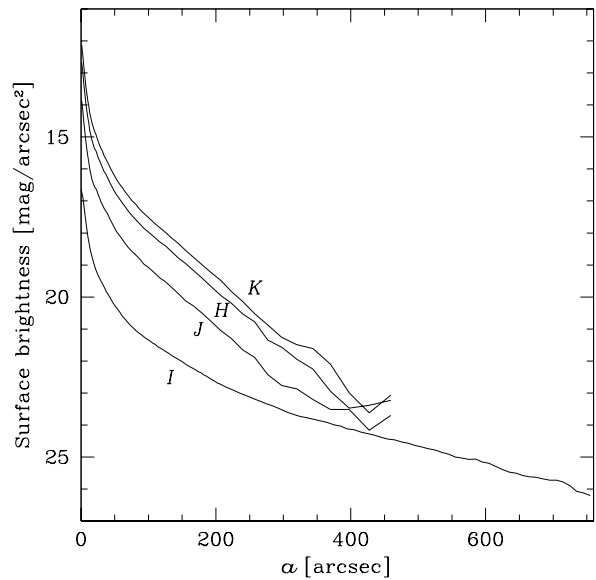}}
\caption{\small
Radial surface brightness profile of Maffei~1 in $IJHK$ along
the galaxy major axis. The $I$-band data are from Buta \&
McCall~(\cite{BMC1}) and the $JHK$ data from the 2MASS Large Galaxy
Atlas (Jarrett et al.~\cite{JCC}). The shifts between the curves
partly reflect the chromatic extinction.}
\label{pro}
\end{figure}

With a distance of only 3 Mpc and an extinction corrected $I$-band
luminosity of $M_I=-22$ magnitude (Fingerhut et al.~(\cite{F}),
Maffei~1 is the nearest normal giant elliptical galaxy to the
Milky~Way. Its apparent size on the sky is about $2/3$ that of the
full moon. Moreover, with Galactic coordinates
$(\ell,b)=(135.9^{\circ},-0.6^{\circ})$, this galaxy lies very close
to the Galactic plane and hence represents an excellent target for a
microlensing search of compact objects in the outer Galactic disc.
Maffei~1 has been fairly well studied in the recent past.
Figure~\ref{obs} shows the near-IR aspect of the galaxy and
Fig.~\ref{pro} its radial profile in several photometric bands. The
large range in surface brightness will show very convenient for
optimising the classical lensing detection rates. The high extinction
by dust inherent to low latitude objects amounts to $A_I=2.64$
magnitude in the optical towards Maffei~1 (Buta \&
McCall~\cite{BMC1}), but is reduced to $A_K=0.57$ magnitude in the
near-IR (Fingerhut et al.~\cite{F}). The equatorial coordinates
$\alpha=02\!:\!36\!:\!35.4$, $\delta=+59\!:\!39\!:\!16.5$ (J2000) mean
that ground-based observations of Maffei~1 have to be done from the
northern hemisphere and that the optimal period is around
October-November. During this period and if observing from Mauna Kea
at a geographic latitude of $+19^{\circ}$, Maffei~1 is about 9 and
10~hours at airmass less than 2.1 and 2.4 respectively.
\par Reported values for the mean isophotal axis ratio and position
angle measured eastward from north are respectively 0.73 and
$83.9^{\circ}$ in the $I$-band (Buta \& McCall~\cite{BMC1}), and 0.79
and $85.5^{\circ}$ in the $K$-band (Jarrett et al.~\cite{JCC}).
Hence the apparent principal axes of Maffei~1 nearly coincide with
the $\alpha$ and $\delta$ coordinate axes.
\par Maffei~1 is part of the Maffei group, which also contains another
large galaxy known as Maffei~2. This is a spiral galaxy which lies
even closer to the Galactic plane, at \hbox{$b=-0.3^{\circ}$}, and
thus provides another possible target which will not be considered
here.

\section{Basic microlensing properties}
\label{basic}

This section shortly reviews the light curves associated to
microlensing events, and evaluates the typical Einstein radius,
microlensing optical depth and timescale for the lensing of
extragalactic sources by H$_2$ cores in the outer Galactic disc.
\begin{figure*}[t!]
\centerline{\includegraphics[width=15.8cm]{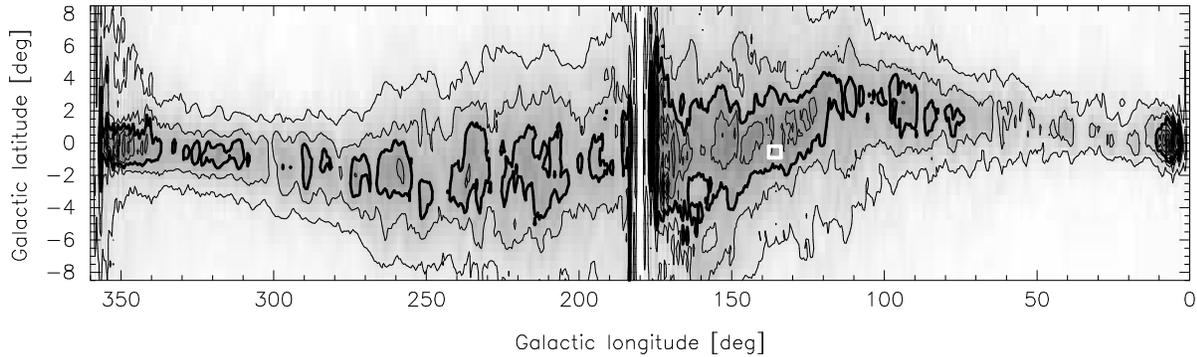}}
\caption{\small
Microlensing optical depth map near the Galactic plane of the
H$_2$ globule cores assuming that their 3D mass density is twice that
of the observed neutral hydrogen, for distant extragalactic sources
($D_{\rm S}\!\gg\! D_{\rm L}$).
The contours are spaced by $\Delta \tau_{-6}=0.2$
and the thick contour stands for $\tau_{-6}=0.6$. The location of
Maffei~1 is indicated by the white square. The HI data are from
Hartmann \& Burton~(\cite{HB}), Burton \& Liszt~(\cite{BL}) and Kerr
et al.~(\cite{KBKJ}).}
\label{tau}
\end{figure*}

\subsection{Light curves}
\label{lc}

Microlensing of a stellar source is the amplification of its measured
flux due to foreground lensing objects passing close to its line of
sight. If $\theta$ is the angular separation of the lens relative to
the source, the magnification factor for a point source is given by:
\begin{equation}
A(x)=\frac{x^2+2}{x\sqrt{x^2+4}},\hspace*{0.5cm}
x\equiv \theta/\theta_{\rm E},
\label{amp}
\end{equation}
where $\theta_{\rm E}\equiv r_{\rm E}/D_{\rm L}$ is the apparent
radius of the Einstein ring (see Sect.~\ref{re}). The motion of the
lens with respect to the line of sight to the source is generally well
approximated by a uniform velocity $\vec{v}$, and thus:
\begin{equation}
x(t;t_{\circ},\beta,\omega)=\sqrt{\omega^2(t-t_{\circ})^2+\beta^2},
\end{equation}
where $t_{\circ}$ is the time of maximum magnification,
$\beta=x_{\rm min}$ the impact parameter in units of $\theta_{\rm E}$
and $\omega^{-1}\equiv r_{\rm E}/v$ the half timescale.
\par As discussed in G96, the truly measurable quantity in crowded
fields is not the magnification itself, but the excess flux
$F(A\!-\!1)$, where $F$ is the unmagnified flux of the lensed
star. And for small impact parameters ($\beta\ll 1$) and near the
maximum magnification ($|\omega(t-t_{\circ})|\ll 1$), this excess flux
takes the limiting form:
\begin{equation}
F[A(t;t_{\circ},\beta,\omega)-1]\rightarrow
\frac{F}{\beta}G(t;,t_{\circ},\omega_{\rm eff}),
\end{equation}
where
\begin{equation} 
G(t;,t_{\circ},\omega_{\rm eff})\equiv \frac{1}
{\sqrt{\omega_{\rm eff}^2(t-t_{\circ})^2+1}}
\end{equation}
and $\omega_{\rm eff}\equiv \omega/\beta$. Hence, in the case where
the intrinsic signal of the source star is well below the background
noise and the event detectable only through high amplification, i.e.
small $\beta$, only $F/\beta$ and $\omega_{\rm eff}$ can be
constrained. This is the so called ``spike'' regime of pixel lensing.
Measuring individually all the $F$, $\beta$ and $\omega$ parameters
requires the wings of the light curve.

\subsection{Einstein radius}
\label{re}

In the case of extragalactic stars lensed by H$_2$ cores in the outer
Galactic disc, the absolute Einstein radius is:
\begin{equation}
r_{\rm E}= 2.3\cdot 10^6\;{\rm km} \left(\frac{m_{\rm L}}{M_{\oplus}}
      \right)^{1/2}\left[\frac{D_{\rm L}}{10\;{\rm kpc}}\cdot
      \frac{D_{\rm S}
      -D_{\rm L}}{D_{\rm S}}\right]^{1/2}
\label{req}
\end{equation}
(or $3.4\;R_{\odot}$), where $m_{\rm L}$ is the mass of the cores,
$M_{\oplus}\!=\!3.0\cdot 10^{-6}\;M_{\odot}$ the Earth mass, and
$D_{\rm S}$ and $D_{\rm L}$ are the distances of the source and the
lens respectively. For sources in the Maffei group,
$(D_{\rm S}-D_{\rm L})/D_{\rm S}\approx 1$. Hence $r_{\rm E}$ is
typically two orders of magnitude larger than the maximum radius of
the H$_2$~cores, ensuring that the cores are sufficiently compact for
the above formula to be applicable. Moreover, at the distance of the
Maffei group, the apparent Einstein radius of the cores represent a
transverse distance
$(D_{\rm S}/D_{\rm L})r_{\rm E}\approx 850\;R_{\odot}$, large enough
to consider even the brightest stars as point sources for not too
small impact parameters. More quantitatively, for sources of finite
apparent radius $\theta_{\rm S}$, Gould~(\cite{G95b}) has shown that
the true magnification {\it exceeds} the point source magnification of
Eq.~(\ref{amp}) by up to $\sim\! 25$\% as long as
$\beta\ga 0.5\theta_{\rm S}/\theta_{\rm E}$. Near the tip of the red
giant branch, the stars have a radius of about $100\;R_{\odot}$, which
means that Eq.~(\ref{amp}) provides a useful lower limit for at least
$\beta\ga 0.06$.

\subsection{Microlensing optical depth}

The microlensing optical depth $\tau$ towards a source star is the
probability that this star lies within the apparent Einstein radius of
any intervening lens, i.e. that its flux is magnified by at least
$\sim\! 0.32$ magnitude. 
\par To estimate the optical depth towards extragalactic sources due
to Galactic H$_2$ cores, we make the following assumptions: (i) most
of the H$_2$ globule mass is contained in the condensed core, (ii) the
space distribution of the globules is similar to that of the HI gas,
as justified in Sect.~1, and (iii) the H$_2$ globules represent twice
the HI mass. This last assumption is based on the fact that the
scatter in the Tully-Fisher relation for external galaxies is
significantly reduced if the contribution of the HI mass is multiplied
by a factor up to 3, and not anymore for larger factors (Pfenniger,
private communication). For the Milky Way, this is a rather
pessimistic assumption if H$_2$ is to make most of the Galactic dark
mass. Indeed, adding 3 times the HI mass distribution to the stellar
mass distribution is insufficient to rise the rotation curve above
190~km\,s$^{-1}$ in the outer disc. This may however compensate to
some extent the optimistic first assumption.
\par Resorting to the available HI data cubes, the traditional
integral over $D_{\rm L}$ that defines the microlensing optical depth
can be converted into a sum over the velocity bins:
\begin{equation}
\tau = \frac{4\pi G\, X m_{\rm p}\, \gamma}{c^2}\cdot
\sum_{\scriptsize \begin{array}{c} i \\
\hspace*{-0.5cm}0\leq D_i\leq D_{\rm S}\hspace*{-0.7cm}\end{array}}
T_{\rm b}(V_i)D_i\frac{D_{\rm S}-D_i}{D_{\rm S}}\Delta V,
\end{equation}
where $T_{\rm b}$ is the brightness temperature, $\Delta V$ the
velocity resolution element, $V_i$ the velocity in the $i$th bin,
$D_i$ the distance corresponding to this velocity, $X$ the conversion
factor of $T_{\rm b}$ into HI atom column density per unit velocity
($X=1.823\cdot 10^{22}$ in MKSA units, see Burton~\cite{B}),
$\gamma=2$ the assumed H$_2$/HI mass ratio, and $m_{\rm p}$ the proton
mass. To infer $D_i$ from $V_i$, a simple kinematic model where the
Sun and the HI gas move on circular orbits with constant circular
velocity $v_{\rm c}(R)=v_{\circ}=200$~km\,s$^{-1}$ and the
galactocentric distance of the Sun $R_{\circ}=8$~kpc is adopted. The
computed optical depths do only marginally depend on the chosen model.
The contribution of the inner Galaxy, complicated by the well known
near and far side distance degeneracy relative to the tangent point,
is not of special interest here and is therefore discarded.
\par The resulting microlensing optical depth map is shown in
Fig.~\ref{tau}. The advantage of our procedure is that it takes fully
account of the deviations of the HI disc from the $b=0$ plane. The
H$_2$-core optical depth can reach values as high as $\tau_{-6}=1$,
where $\tau_{-6}$ is the optical depth expressed in units of
$10^{-6}$. This is nearly as much as the values of $\sim\! 2$ measured
by the microlensing experiments towards the Galactic bulge (e.g.
Popowski et al.~\cite{PVG}). In the direction of Maffei~1,
$\tau_{-6}\approx 0.7$, very close to these maximum values. Note that
$\tau$ does depend neither on $m_{\rm L}$ for compact lenses, nor on
the lens velocity distribution.

\subsection{Timescale}
\label{times}

With the same kinematic model as used in the former section, assuming
that the Sun and the H$_2$ globules are on circular orbits and
neglecting the proper motion of the Maffei~1 stars relative to the
Galactic centre, the transverse velocity of the globules relative to
these stars line of sight is:
\begin{equation}
v=[
\sqrt{1-\frac{\sin^2{\ell}}
{1-2D_{\rm L}/R_{\circ}\cos{\ell}+(D_{\rm L}/R_{\circ})^2}}+\cos{\ell}]
\cdot v_{\circ},
\label{vm}
\end{equation}
where $\ell$ is the Galactic longitude of Maffei~1. In particular, for
$D_{\rm L}=10$~kpc, one gets $v=0.225v_{\circ}=45$~km\,s$^{-1}$. The
Earth's motion relative to the Sun adds a quasi constant velocity of
about $17$~km\,s$^{-1}$ in October-November, which is around the
yearly maximum value, and the Sun's motion relative to the LSR about
$-11$~km\,s$^{-1}$, implying a resulting $v\approx 50$~km\,s$^{-1}$.
At such a transverse velocity, the timescale of a microlensing event,
corresponding to the crossing time of the Einstein ring, amounts to:
\begin{equation}
2\omega^{-1} \equiv \frac{2r_{\rm E}}{v}=26.0\;{\rm hours}
\left(\frac{r_{\rm E}}{3.4\;R_{\odot}}\right)
\left(\frac{50\;{\rm km/s}}{v}\right).
\label{ts}
\end{equation}
Hence one observing night enables to measure at least the third of a
lensing light curve if $m_{\rm L}\la 1\;M_{\oplus}$, but it may not be
possible to get the full curve from a single ground-based telescope.
According to Eq.~(\ref{req}), $\omega^{-1}\propto\sqrt{m_{\rm L}}$,
and thus the timescale decreases with lens mass.
\par Equation~(\ref{req}) also leads to
$\omega^{-1}\!\propto\!\sqrt{D_{\rm L}}/v$ at fixed lens mass, which
for the simple mean velocity model expressed in Eq.~(\ref{vm}),
implies that the timescale is constant to within about 10\% in the
distance range $2\;{\rm kpc}\la D_{\rm L}\la10\;{\rm kpc}$ and
significantly increases only at shorter distances.
\begin{table}
\centering
\caption{\small
Parameters of the adopted $I$- and $K$-band luminosity
functions, and the resulting fluctuation magnitudes.}
\begin{tabular}{crc} \hline\hline \vspace*{-.35cm} \\
 & \multicolumn{2}{c}{~~Filter} \\
 & $I$~ & ~~~$K$ \\
\vspace*{-.35cm} \\ \hline
$n$           &  $2.0$ & ~$1.675$ \\
$M_{\rm lim}$ & $-4.1$ &  $-7.5$ \\
$M_{\circ}$   &  $3.0$ &~~~~$\infty$ \vspace*{0.1cm}\\
$\bar{M}$     & $-2.1$ &  $-6.0$ \\ \hline
\label{LF}
\end{tabular}
\end{table}

\section{Luminosity function and fluctuation magnitude}
\label{lum}

In order to calculate detection rates, we will need to quantify the
surface brightness noise due to the finite number of stars in a
galaxy, and specify a realistic stellar luminosity function $\Phi(F)$
applicable to the Maffei~1 galaxy.
\par The surface brightness noise is conveniently characterised by the
``fluctuation flux'', which is defined as the ratio of the second to
the first moments of the luminosity function:
\begin{equation}
F_*\equiv \frac{\int_0^{\infty}\Phi(F)F^2{\rm d}F}
{\int_0^{\infty}\Phi(F)F{\rm d}F}.
\end{equation}
This flux, which is also fundamental to the surface brightness
fluctuation method for measuring distances, represents the flux of the
stars in a system made of identical stars that would statistically
produce the same surface brightness profile and the same surface
brightness fluctuation as the real stellar system.
\par The observed absolute fluctuation magnitudes of galaxies in $IHK$
as a function of intrinsic colour are:
\begin{eqnarray}
&& \hspace*{-0.3cm}\bar{M}_{\!I}=(-1.74\pm 0.07)+(4.5\pm 0.25)
[(V\!-\!I)_{\circ}-1.15]\\
&& \hspace*{4.8cm}\hbox{(Tonry et al.~\cite{TBA}),}\nonumber \\
&& \hspace*{-0.3cm}\bar{M}_{\!H}=(-5.03\pm 0.03)+(5.1\pm 0.5)
[(V\!-\!I)_{\circ}-1.16]\\
&& \hspace*{4.8cm}\hbox{(Jensen et al.~\cite{JTB}),}\nonumber \\
&& \hspace*{-0.3cm}\bar{M}_{\!K_{\rm s}}\!=(-5.84\pm 0.04)+(3.6\pm 0.8)
[(V\!-\!I_{\rm C})_{\circ}-1.15]\\
&& \hspace*{4.8cm}\hbox{(Liu et al.~\cite{LGC}).}\nonumber
\end{eqnarray}
Fingerhut et al.~(\cite{F}) report $(V\!-\!I)_{\circ}=1.06\pm 0.09$
for Maffei~1, which implies $\bar{M}_{\!I}\!=\!-2.1$ and
$\bar{M}_{\!K_{\rm s}}\!=\!-6.2$ for this galaxy. The agreement with
the theoretical values derived from population synthesis models is
very good in the $I$-band, but not in the $K$-band, where the models
predict $\bar{M}_{\!K}\!=\!-4.9$ (Cantiello et al.~\cite{CRB}). The
fact is that the fluctuation magnitude is much brighter in the near-IR
than in the optical.
\par The stellar luminosity functions of nearby galaxies show a power
law behaviour at the bright end over a large range of magnitude (apart
from the red clump bump), with a rather sharp cut-off at the tip of
the red giant branch and a flattening at fainter magnitudes. In this
paper, we will use the following functional form:
\begin{equation}
\begin{array}{l}
\Phi(F) =
\left\{\begin{array}{cl}
       {\displaystyle 0} & F > F_{\rm lim},\\
       {\displaystyle \frac{C}{F^n}} & F_{\circ} \leq F \leq F_{\rm lim},\\
       {\displaystyle \hbox{\tiny ${\cal W}$}\frac{C}{F_{\circ}^n}} &
       F \leq F_{\circ},
       \end{array} \right.
\end{array}
\label{B3}
\end{equation}
where the variable $F$ is the observed flux, $F_{\rm lim}$ the bright
end cut-off flux, $F_{\circ}$ the ``knee'' flux where the power law
breaks and $\Phi(F)$ flattens, and {\tiny ${\cal W}$} a parameter to
tune the contribution of faint stars. Our results will only marginally
depend on the precise choice of this last parameter and we therefore
arbitrarily set $\hbox{\tiny ${\cal W}$}=0$. If $n\neq 1$, the
normalisation factor is:
\begin{equation}
C=\frac{(n-1)(F_{\rm lim}F_{\circ})^{n-1}}
{[1+(n-1)\hbox{\tiny ${\cal W}$}]F_{\rm lim}^{n-1}-F_{\circ}^{n-1}}.
\end{equation}
In the realistic case $F_{\circ}\ll F_{\rm lim}$, the resulting
absolute fluctuation magnitude is well approximated by:
\begin{equation}
\begin{array}{l}
\bar{M} =
\left\{\begin{array}{ll}
       {\displaystyle \!M_{\rm lim}+2.5\log{[\nu(M_{\circ}\!-\!M_{\rm lim})
       +\frac{\hbox{\tiny ${\cal W}$}}{2}]}} & n=2,\\
       {\displaystyle \!M_{\rm lim}+2.5\log{\left(\frac{3-n}{2-n}\right)}}
              & n<2\!-\!\epsilon,\hspace*{-1cm}
       \end{array} \right.
\end{array}
\label{MS}
\end{equation}
where the subscripted absolute magnitudes correspond to the similarly
subscripted fluxes, $\nu=0.4\ln{10}$, and $\epsilon$ is a small
non-zero positive number introduced because the given approximation
fails when $n\rightarrow 2$.
\par The parameters of $\Phi(F)$ are determined as follow, assuming
that Maffei~1 and the bulge of the Milky~Way have similar luminosity
functions. In the $I$-band,  the slope of the observed magnitude
function in the Galactic bulge is
$s\equiv {\rm d}\log_{10}\Phi(m)/{\rm d}m=0.4$ for the bright stars
(Terndrup et al.~\cite{TFW}), thus $n=2.5\cdot s+1=2$. Note that this
value differs from $n=1$ assumed in G96. Here the term $+1$ arises
because ${\rm d}F/{\rm d}m\propto F$. The tip of the red giant branch
lies at $M_{\rm lim}\approx -4.1$ (e.g. Karachentsev et
al.~\cite{KSD}) and the knee magnitude at $M_{\circ}\approx 3.0$
(Holtzman et al.~\cite{HWB}), and for these values Eq.~(\ref{MS})
yields \hbox{$\bar{M}=-2.1$}, in perfect agreement with the expected
fluctuation magnitude of Maffei~1. The agreement would be slightly
worse for $\hbox{\tiny ${\cal W}$}=1$, probably because our luminosity
function ignores supergiants brighter than $M_{\rm lim}$.
\par In the $K$-band, the observed bright-end slope in the Galactic
bulge is $s=0.27\pm 0.03$ down to the faintest detected stars, leading
to $n=1.675$, and $M_{\rm lim}\approx -7.5$ (Tiede et al.~\cite{TFT}).
For such a value of $n$, Eq.~(\ref{MS}) predicts no significant
dependence of $\bar{M}$ on $M_{\circ}$, so that we arbitrarily chose
$M_{\circ}=\infty$, and one gets $\bar{M}=-6.0$, consistent with the
quoted value based on observations. Note that $\Phi(F)$ is not
normalisable for this choice of $M_{\circ}$, but $\bar{M}$ is
nevertheless well defined. Table~\ref{LF} lists the retained values of
the parameters.
\par Finally, we will also need to know the number of stars per unit
solid angle and per unit galaxy surface brightness with flux larger
than $F$, which for our luminosity function and for
$F_{\circ}\leq F\leq F_{\rm lim}$, if $F_{\circ}\ll F_{\rm lim}$ and
$1<n<3\!-\!\epsilon$, is:
\begin{equation}
N(F)=\frac{1}{(n-1)(3-n)}\frac{F_*}{F_{\rm lim}^2}
\left[\left(\frac{F_{\rm lim}}{F}\right)^{n-1}-1\right],
\label{cum}
\end{equation}
where $\epsilon$ is as above.

\section{Detection rates}
\label{rate}

In this section, we will derive estimates of the expected detection
rates of the H$_2$-core microlensing events when targeting the
Maffei~1 galaxy with existing observing facilities, assuming
$\tau_{-6}=0.7$ and $\omega^{-1}=13$ hours, as estimated in
Sect.~\ref{basic}, and a homogeneous stellar population with $\Phi(F)$
as detailed in Sect.~\ref{lum}. Both classical lensing and pixel
lensing approaches are investigated, and the results are presented for
one optical and one near-IR filter, i.e. for the $I$- and
\hbox{$K$-bands}.

\subsection{Classical lensing}

Classical lensing is the microlensing of sources that are resolved
even when not magnified. In this case the source counts contribute
significantly to the noise and the signal-to-noise ratio of a star in
a given exposure is:
\begin{eqnarray}
Q\!=\!\frac{(\alpha F)t_{\rm exp}}{\sqrt{(\alpha F)t_{\rm exp}
\!+\!\Omega_{\rm psf}\{(\alpha S)[1\!+\!g(\alpha F_*)t_{\rm exp}]
\!+\!(\alpha S_{\rm sky})\}t_{\rm exp}}}\hspace*{-1cm}&&\nonumber\\
\label{Q1}
\end{eqnarray}
where $F$ is the flux from the star, $S$ the galaxy surface
brightness, $S_{\rm sky}$ the sky surface brightness including the
instrument background, $t_{\rm exp}$ the exposure time,
$\Omega_{\rm psf}$ the solid angle covered by the point spread
function (PSF), $F_*$ the galaxy fluctuation flux, and $\alpha$ the
number of electrons counted by the detector per unit time and unit
flux. If $m$ is the apparent magnitude of the star and $Z$ the
detector zeropoint, i.e. the magnitude of a star producing one
electron per second, then $\alpha F=10^{0.4(Z-m)}$.
\par The term $\Omega_{\rm psf}(\alpha S)(\alpha F_*)t_{\rm exp}^2$
represents the fluctuation flux contribution to the squared noise if
the background stars were seen as point-like, and the factor $g$ is a
correction taking into account the finite seeing:
\begin{equation}
g\equiv \frac{1}{\Omega_{\rm psf}}\int[\int_{\Omega_{\rm psf}}
f(\vec{x}'-\vec{x}){\rm d}\vec{x}']^2{\rm d}\vec{x},
\end{equation}
where $f(\vec{x})$ is the PSF. We will consider
$\Omega_{\rm psf}=\pi\,\theta_{\rm see}^{\,2}$, with the seeing
$\theta_{\rm see}$ defined as the PSF full width at half maximum. For
a Gaussian PSF with standard deviation~$\sigma$,
$\theta_{\rm see}=2\sqrt{\ln{4}}\sigma$ and $g=0.544$.
\par Equation~(\ref{Q1}) can easily be inverted to get the minimum
flux $F_{\rm min}(S,t_{\rm exp})$ necessary to detect a star with a
given signal-to-noise threshold $Q>Q_{\rm min}$. This flux is of
course a function of the exposure time, but cannot reach arbitrarily
small values as $t_{\rm exp}\rightarrow \infty$. The fluctuation noise
indeed imposes the lower limit:
\begin{equation}
F_{\rm min}(S,\infty)=
Q_{\rm min}\sqrt{g\Omega_{\rm psf}(\alpha S)(\alpha F_*)}.
\label{lim}
\end{equation}
The total number of resolvable stars per unit solid angle then amounts
to:
\begin{equation}
N_{\rm res}(S,t_{\rm exp})=S\,N(F_{\rm min}),
\label{nres}
\end{equation}
where $N(F)$ is given by Eq.~(\ref{cum}). Now the classical event rate
for one star is:
\begin{equation}
\Gamma_{\circ}=\frac{2}{\pi}\omega\tau,
\label{Go}
\end{equation}
thus the number of lensing events per unit solid angle with peak
magnification within an observing time interval $\Delta t$ will be:
\begin{equation}
N_{\rm det}=N_{\rm res}\Gamma_{\circ}\Delta t
=\frac{2}{\pi}\tau\,S\,N(F_{\rm min})\,\omega\Delta t.
\end{equation}
\par
\begin{figure}
\centerline{\includegraphics[width=8cm]{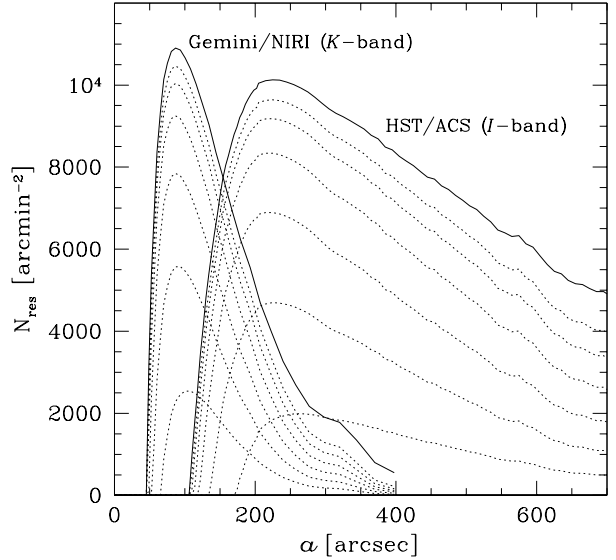}}
\caption{\small
Expected number of resolvable stars per unit solid angle in
Maffei~1 as a function of angular radius $a$ along the major axis of
the galaxy and exposure time, for two combinations of telescopes and
detectors, including adaptive optics in the $K$-band, and using the
surface brightness profiles plotted in Fig.~\ref{pro}. The dotted
lines are for $t_{\rm exp}=1,2,4,8,16,32$~sec for Gemini/NIRI, and for
$t_{\rm exp}=1.25,2.5,5,10,20,40$~min for HST/ACS, and the solid lines
represent the upper limits resulting from Eq.~(\ref{lim}). The
detection threshold is set to 3~sigmas. The adopted telescope
zeropoints, sky backgrounds and seeing angles are listed in
Table~\ref{pixres}.}
\label{classic}
\end{figure}

 Figure~\ref{classic} displays $N_{\rm res}$ for Maffei~1 as a
function of distance from the galaxy centre when observations are
carried out in the \hbox{$I$-band} with the ACS camera on the Hubble
Space Telescope (HST), and in the $K$-band with the NIRI plus adaptive
optic instruments on the Gemini North telescope, assuming
$Q_{\rm min}=3$. Extinction is taken into account, with the values of
$A_I$ and $A_K$ mentioned in Sect.~\ref{maf}. The choice of
$Q_{\rm min}$ is rather small, but since the same field is observed
several times, one can stack the exposures to enhance the
signal-to-noise ratio and hence avoid false star detections.
\par The number of resolvable stars increases linearly at low
$t_{\rm exp}$ and saturates at large $t_{\rm exp}$ due to the minimum
flux limit provided by Eq.~(\ref{lim}). The exposure time at the
transition between the two regimes is around 10 min for HST/ACS and
only 5~sec for Gemini/NIRI. For longer exposures, doubling
$t_{\rm exp}$ does not efficiently increase $N_{\rm res}$ anymore. The
shorter integration time required with NIRI comes from the higher
values of $S$ and $F_*$, and the lower extinction in~$K$, even if the
sky is much brighter in this band than in the $I$-band.
\par In both the HST/ACS and Gemini/NIRI cases, there exists an
optimal radius which maximises the number of resolvable stars. Near
the centre, the fluctuation noise is so high that
$F_{\rm min}\!>\!F_{\rm lim}$, preventing the detection of individual
stars. This is consistent with the fact that no stars are resolved in
Buta \& McCall's~(\cite{BMC2}) HST data of the central region of
Maffei~1. At increasing distance, $F_{\rm min}$ decreases below
$F_{\rm lim}$ and $N_{\rm res}$ begins to rise abruptly. But at even
larger distance, this trend reverses due to the decline with $S$ of
the stellar number density. The radius of maximum $N_{\rm res}$ does
not vary much with exposure time, except at the short exposure end,
where it increases with decreasing $t_{\rm exp}$, and is smaller in
$K$ than in $I$ mainly because of the strong chromatic dependence of
the luminosity function.
\par For HST/ACS, the optimum distance along the major axis is around
230~arcsec from the centre. At this radius, if $\Omega$ is the solid
angle covered by the detector (see Table~\ref{pixres}) and
$t_{\rm exp}=10$~min, one should be able to resolve as much as
$\Omega N_{\rm res}\approx 90\,000$ stars, yielding
$N_{\rm det}\approx 0.003(\Delta t /{\rm hour})$. This is a rather low
detection rate that would require many days of HST time if one
restricts to a single field. However, thanks to the large apparent
size of Maffei~1 and as suggested in G96, the situation can be
improved when observing several fields simultaneously. If we require
at least 8 exposures per field during the typical half lensing
timescale of 13 hours, this implies one exposure per field every
96~hours of HST orbital time. Given the 56 hours of visibility period
for Maffei~1, the 10.1 minutes of overhead time per exposure\footnote{
Unfortunately, the ACS field of view is too large to keep the same
guide star when pointing toward adjacent fields; the overhead would be
substantially less otherwise.} and the minimum exposure time of
339~sec to avoid additional buffer dump overheads (see Pavlovsky et
al.~\cite{PAL}), this means that three different fields can be
observed per orbit, with $t_{\rm exp}=8.5$~min at most.
Figure~\ref{field} displays four optimal possibilities where to choose
such fields. The detection rate can therefore be tripled to finally
get one lensing event every 5 days. Taking the risk of missing short
events, one could also reduce the time sampling and measure 6 fields
in periods of two HST orbits. Note that if the entire surface of
Maffei~1 could be exploited, there would be over $2\cdot 10^6$ useful
stars.
\par For Gemini/NIRI, the total number of resolvable stars in Maffei~1
is $\int N_{\rm res}{\rm d}\Omega\approx 4\cdot 10^5$. Hence even if
the very short $t_{\rm exp}$ would allow to mosaic the full area
covered by these stars, and assuming 10 hours of observability per
night, the detection rate would be only of one event every 8 days,
somewhat less than with HST.
\par In classical lensing, the effective seeing is of crucial
importance because it directly determines the number of resolvable
stars. For instance, without adaptive optics and a seeing of $0.4''$,
NIRI would resolve only about $2\cdot 10^4$ stars over all Maffei~1 at
$Q_{\rm min}=3$. Furthermore, NICMOS on HST, with its $K$-band
diffraction limited effective seeing of $0.2''$, would see about
$6\cdot 10^4$ stars at the same $Q_{\rm min}$ and could monitor only a
tiny fraction of them. Hence, while high sensitivity does not really
matter in the \hbox{$K$-band} here, large telescopes are nevertheless
essential to alleviate the diffraction limit. Likewise, ground-based
optical observations, which do not yet benefit from adaptive optic
corrections, are hopeless at the moment for classical lensing
searches.
\par However, one advantage of classical lensing over pixel lensing is
that it does not suffer from non-photon noise, like variable PSF,
misalignment and pixelisation noise. In principle, classical lensing
also always give access to the event timescales. To check for the
non-chromatic dependence of the light curves would require
observations with a second filter, which has been ignored here.

\subsection{Pixel lensing}
\label{pl}

In pixel lensing, the source is not resolved and the noise is
dominated by the stellar, sky plus instrumental background. The
lensing events are revealed by subtracting a reference image ideally
with no lensing event to the images with the ongoing events. The
resulting difference images then show a PSF of the excess flux due to
the lensing magnification. Neglecting the contribution of the excess
flux to the noise, the signal-to-noise ratio in a difference image
obtained from an exposure at time $t_i$ of an event with maximum
magnification at time $t_{\circ}$ is:
\begin{equation}
Q_i=\frac{(\alpha F)[A(t_i;t_{\circ},\beta,\omega)-1]t_{\rm exp}}
{\sqrt{\mu\Omega_{\rm psf}[(\alpha S)+(\alpha S_{\rm sky})]t_{\rm exp}}},
\end{equation}
where $t_{\rm exp}$ is the exposure time, $F$ the unamplified flux of
the lensed star, $\mu$ a factor that takes into account the noise
in the reference image, and $S$, $S_{\rm sky}$, $\Omega_{\rm psf}$ and
$\alpha$ are as for Eq.~(\ref{Q1}). Note that in the absence of pixel
noise, the subtraction performed when creating the difference image
removes the fluctuation noise.
\begin{figure}
\centerline{\includegraphics[width=8cm]{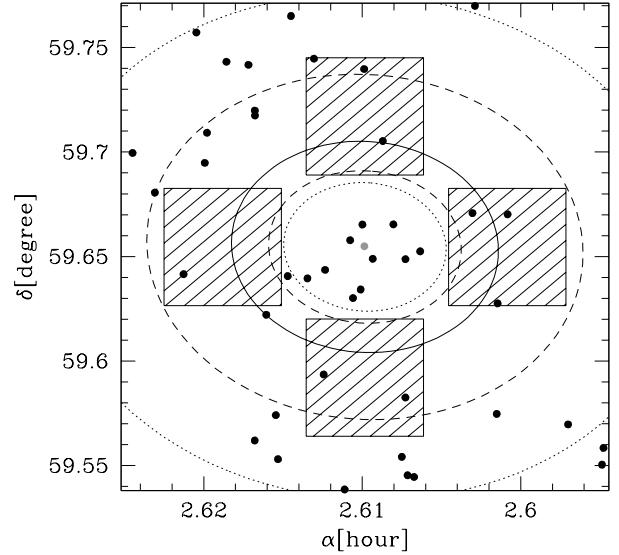}}
\caption{\small
Location of four optimal HST/ACS fields in Maffei~1 for the
search of classical lensing events. The full, dashed and dotted lines
represent the isophotes where the $I$-band surface density of
resolvable stars $N_{\rm res}$ is maximum, $80\%$ of maximum, and
$50\%$ of maximum respectively. The dots represent bright guide stars
for near-IR adaptive optic observations, obtained from the STSI Guide
Star Catalogue I.}
\label{field}
\end{figure}
\par Now let's assume that the event is observed through $N$ exposures
from time $t_1$ to time $t_2$, a time interval not necessary covering
the entire event. The total signal-to-noise ratio of the measured
portion of the event then is $Q=(\sum_{i=1}^N Q_i^2)^{1/2}$. In
practice, an event is searched for by summing the difference images
weighted by all possible and properly normalised $(A-1)$ factors,
intervening as a ``filter'' function. When the parameters of this
factor coincide with those of the event, the signal-to-noise ratio of
the event PSF on the summed image reaches a maximum with value given
precisely by $Q$. If the exposures are taken uniformly at time
intervals small relative to $\omega_{\rm eff}^{-1}$, and that the sky
noise and the PSF remain constant with time, the sum in $Q$ can be
converted into an integral yielding:
\begin{equation}
Q^2=\frac{\pi\eta(\alpha F)^2}{\mu\Omega_{\rm psf}
[(\alpha S)+(\alpha S_{\rm sky})]\omega}
\frac{\zeta(t_{\circ},\beta,\omega;t_1,t_2)}{\beta},
\label{tot}
\end{equation}
\begin{eqnarray}
\zeta(t_{\circ},\beta,\omega;t_1,t_2) & \equiv &
       \frac{\int_{t_1}^{t_2}[A(t;t_{\circ},\beta,\omega)-1]^2{\rm d}t}
       {\beta^{-2}\int_{-\infty}^{\infty}
       G(t;t_{\circ},\omega_{\rm eff})^2{\rm d}t}\nonumber\\
& \hspace*{-4cm} = & \hspace*{-2.0cm}
  \frac{\omega\beta}{\pi}
  \left[2(1-\frac{1}{\beta}\sqrt{\beta^2+4-\frac{4}{1+\frac{\beta^2}
  {\omega^2t^2}}})t+\frac{1}{\omega\beta}\cdot \right. \nonumber\\
& & \hspace*{-2.0cm}
      \{\arctan{(\frac{\omega t}{\beta})}-\frac{\beta}
      {\sqrt{\beta^2+4}}\arctan{(\frac{\omega t}{\sqrt{\beta^2+4}})}
  \nonumber\\
& & \hspace*{-2.0cm}
    +2\beta[\sqrt{\beta^2+4}\,E(\frac{\omega t}{\sqrt{\omega^2t^2+\beta^2}},
      \frac{2}{\sqrt{\beta^2+4}}) \nonumber\\
& & \hspace*{-2.0cm}\left.
      -\frac{\beta^2+2}{\sqrt{\beta^2+4}}\,
      F(\frac{\omega t}{\sqrt{\omega^2t^2+\beta^2}},
      \frac{2}{\sqrt{\beta^2+4}})]\}\right]_{t_1-t_{\circ}}^{t_2-t_{\circ}}
\end{eqnarray}
where $\eta\equiv Nt_{\rm exp}/(t_2-t_1)$ is the time fraction really
dedicated to observe the source galaxy, and $F(z,k)$ and $E(z,k)$ are
the incomplete elliptical integrals of the first and second kind
respectively. The function $\zeta$ is always $\leq 1$. In the case
where $t_1\rightarrow -\infty$ and $t_2\rightarrow \infty$, it depends
only on $\beta$ and coincides with the monotonically decreasing
suppression function defined in G96, with $\zeta\rightarrow 1$ for
$\beta\ll 1$.
\par Only the lensing events above a given signal-to-noise threshold,
say $Q>Q_{\rm min}$, are detectable. For given $F$, $\omega$ and
$t_{\circ}$, this condition and Eq.~(\ref{tot}) define a maximum
impact parameter $\beta_{\rm max}(F,t_{\circ},\omega;t_1,t_2)$ as the
solution for $\beta$ of:
\begin{equation}
\frac{\beta}{\beta_{\rm max}^{\circ}(F)}=
\zeta(t_{\circ},\beta,\omega;t_1,t_2),
\label{Bmax}
\end{equation}
\begin{equation}
\hspace*{0.5cm}\beta_{\rm max}^{\circ}(F)=\frac{\pi\eta(\alpha F)^2}
{Q_{\rm min}^2\mu\Omega_{\rm psf}[(\alpha S)+(\alpha S_{\rm sky})]\omega},
\label{Bmax0}
\end{equation}
such that events with larger impact parameters will not be detected.
As mentioned in Sect.~\ref{re}, there also exists a lower limit for
the impact parameter owing to the finite radius of the stellar sources,
below which the magnification is less than predicted by Eq.~(\ref{amp}).
For simplicity, we will assume that there is a one-to-one relation
between the stellar radius $R_{\rm S}$ and the flux, and in particular
that $R_{\rm S}$ does not depend on colour. A justification to this
assumption is that elliptical galaxies like Maffei~1 essentially
contain old stellar populations and thus the bright end of their
stellar luminosity function is dominated by red giants spanning a
rather small colour range. From Stefan-Bolzmann's law it then follows
that $R_{\rm S}=R_{\rm lim}10^{-(M-M_{\rm lim})/5}$, where $M$ is the
absolute magnitude of the star and
$R_{\rm lim}\approx 100\;R_{\odot}$~the stellar radius at the tip of
the red giant branch. In this form, our approximation is most accurate
for the brightest stars, which are the main sources contributing to
observable lensing events, and will overestimates the derived minimum
impact parameter for fainter stars. Keeping only events with
$\beta\geq 0.5\theta_{\rm S}/\theta_{\rm E}$ thus implies:
\begin{equation}
\beta_{\rm min}(F)=\frac{1}{2}\frac{D_{\rm L}}{r_{\rm E}}
\frac{R_{\rm lim}10^{-M/5}}{D_{\rm S}}.
\label{Bmin}
\end{equation}
The number of detectable events from a star with peak magnification
between times $t_{\circ}$ and $t_{\circ}+{\rm d}t_{\circ}$ then simply
is $\Delta\beta\,\Gamma_{\circ}{\rm d}t_{\circ}$, where:
\begin{equation}
\Delta \beta(F,t_{\circ},\omega;t_1,t_2)
\equiv {\rm max}({\rm min}(\beta_{\rm max},1)-\beta_{\rm min},0)
\label{db}
\end{equation}
and $\Gamma_{\circ}$ is the classical event rate given by
Eq.~(\ref{Go}). The upper limit of~$1$ is introduced in Eq.~(\ref{db})
to exclude events with $\beta>1$, which by definition do not
contribute to the microlensing optical depth.
\par Finally, the total number of detectable events within the
observing time from $t_1$ to $t_2$ and per unit solid angle is
obtained by multiplication with the stellar surface density
$S/\int_0^{\infty}\Phi(F)F{\rm d}F$ and by integration over the
luminosity function and over the magnification peak times:
\begin{equation}
N_{\rm det}=\frac{2}{Q_{\rm min}^2}\frac{\eta}{\mu}\cdot\tau\,
\frac{(\alpha F^*)}{\Omega_{\rm psf}}
\frac{(\alpha S)}{(\alpha S)+(\alpha S_{\rm sky})}
\int_{-\infty}^{\infty}\xi\,{\rm d}t_{\circ},
\label{eqpix}
\end{equation}
\begin{equation}
\hspace*{0.5cm}
\xi(t_{\circ},\omega;t_1,t_2)=
\frac{\int_0^{\infty}\Phi(F)\Delta\beta(F,t_{\circ},\omega;t_1,t_2){\rm d}F}
{\int_0^{\infty}\Phi(F)\beta_{\rm max}^{\circ}(F){\rm d}F}.
\label{xi}
\end{equation}
\par In addition to this detection number, it is also useful to know
what fraction of the detected events will have a measurable timescale.
For this purpose, we use the statistical procedure describe in
Gould~(\cite{G95a}). The rate of electrons produced on the detector by
a lensed source star follows the distribution:
\begin{equation}
{\cal F}(t;F,t_{\circ},\beta,\omega)=
\eta(\alpha F)[A(t,t_{\circ},\beta,\omega)-1] + B,
\end{equation}
where $B=\eta\{(\alpha F)+\Omega_{\rm psf}[(\alpha S)+(\alpha
S_{\rm sky})]\}$ is the background rate including the unmagnified flux
from the star. This background can be accurately determined from data
taken outside the lensing event, like the reference images. Noting
$\lambda_i$, $i=1,..,4$, the free parameters $F$, $t_{\circ}$, $\beta$
and $\omega$, the covariance matrix of these parameters can be
calculated as:
\begin{equation}
{\rm cov}(\lambda_i,\lambda_j)=\{b^{-1}\}_{ij},
\end{equation}
\begin{equation}
\hspace*{0.5cm} b_{ij}\equiv
\int_{t_1}^{t_2}{\cal F}(t;F,t_{\circ},\beta,\omega)
\frac{\partial\ln{\cal F}}{\partial\lambda_i}
\frac{\partial\ln{\cal F}}{\partial\lambda_j}{\rm d}t,
\end{equation}
and thus $\delta\omega^2={\rm cov}(\omega,\omega)$. At each apparent
radius of~the source galaxy, the sub-number
$\Delta N_{\rm det}(\delta w/w\leq\varepsilon)$ of events with
$\delta w/w\leq \varepsilon$ is then computed the same way as for
$N_{\rm det}$, except that $\beta_{\rm max}$ is replaced by the impact
parameter where $\delta w/w=\varepsilon$.
\par We will now estimate $N_{\rm det}$ for observations made with a
selection of telescope and instrument combinations within a period
$\Delta t=t_2-t_1=10$~hours, i.e. the typical duration of a
ground-based observing night, assuming the conditions and properties
listed in Table~\ref{pixres}. The observing overheads, i.e.
$(1-\eta)$, are estimated assuming that a single field is observed and
hence no repeated telescope pointings and guide star acquisitions are
needed. For HST/ACS, in addition to the 40~min of non-visibility time
per orbit, the overheads then amount to
$9.7+2.5\cdot ({\cal N}\!-\!1)$ minutes per orbit plus 0.4~min for the
first orbit, where ${\cal N}$ is the number of exposures per orbit
(see Pavlovsky et al.~\cite{PAL}). This number is constrained by the
detector saturation, which for ACS and the central surface brightness
of Maffei~1 is reached after $\sim\! 20$~min
integration\footnote{Taking a detector gain larger than the default
value of 1 would allow a longer integration time.}. To ensure some
margin for high lensing amplification, we choose ${\cal N}=4$ and
$t_{\rm exp}=9.7$~min, representing a total overhead of about $60\%$.
For Gemini/NIRI, the overhead is about $25\%$ according to the
instrument web pages, and the same overhead is taken for all other
ground-based means. No time is reserved for sky frames and chopping,
as the effect of variable sky and fringes is not expected to hide and
significantly perturb the photometry of the PSFs on the difference
images, and for exposures with other filters. The reference images are
assumed to be obtained from a similar set of observations taken
another night, preferably at a few days interval to completely erase
the trace of presumable events, so that $\mu=2$ (see also Sect.~9.6 in
G96). Although this procedure doubles the required observing time, it
also doubles the number of detectable events since the role of the
event and reference images can be interchanged.
\par
\begin{figure}
\centerline{\includegraphics[width=8cm]{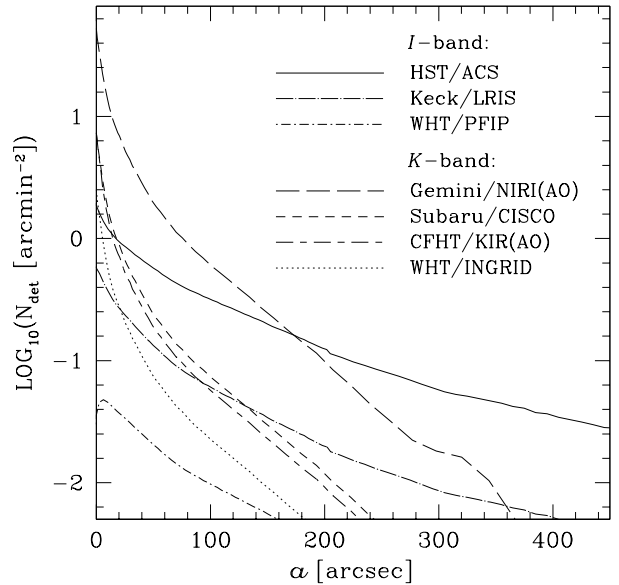}}
\caption{\small
Expected number of detectable pixel lensing events per
arcmin$^2$ in Maffei~1 as a function of angular radius $a$ along the
galaxy major axis, when observing 10 hours with various combinations
of telescopes and detectors, and for the surface brightness profiles
displayed in Fig.~\ref{pro}. The \hbox{signal-to-noise} threshold is
$Q_{\rm min}=5$, and AO indicates the use of adaptive optics. The
observing parameters are specified in Table~\ref{pixres}.}
\label{pixel}
\end{figure}

 Figure~\ref{pixel} plots the resulting radial profiles of
$N_{\rm det}$ for $Q_{\rm min}=5$. The first thing to notice is that,
contrary to classical lensing, all the curves peak near the
centre. Since the non-integral part of $N_{\rm det}$ in
Eq.~(\ref{eqpix}) is also directly proportional to $\eta$ and the
curves rapidly decline with radius, it is therefore best to rely on a
single field at the centre of Maffei~1, hence justifying our overhead
calculations. Moreover, the size of the field of view is not as
limiting as in classical lensing, because the dominant contribution to
the event detection rates comes from the innermost regions.
\par A second point is that, again unlike classical lensing, adaptive
optics (AO) is not an absolute prerequisite for pixel lensing
searches. As an example in the $K$-band, Subaru/CISCO without AO can
detect more events (over the entire detector {\it and} per unit area)
than CFHT/KIR with AO. In particular, this means that ground-based
optical observations can also efficiently chase pixel lensing events.
In fact, in the point source limit ($\beta_{\rm min}=0$) and for fully
measured light curves ($t_1\rightarrow -\infty$ and
$t_2\rightarrow+\infty$), the detection rate integrated over the solid
angle $\Omega$ covered by the detector {\it increases} with increasing
effective seeing if the focal length of the telescope is adjusted in a
way such that $\Omega$ scales as $\Omega_{\rm psf}$ and if
$N_{\rm det}$ remains uniform within the field of view. Indeed, under
these assumptions, $\xi$ gets independent of $t_{\circ}$ and $\omega$
and $\int [N_{\rm det}/\Delta t]{\rm d}\Omega\propto
\Omega/\Omega_{\rm psf}\xi$, and since the suppression function
$\zeta$ reduces to a monotonically decreasing function of $\beta$,
Eqs.~(\ref{Bmax}) and~(\ref{Bmax0}) imply that
$\beta_{\rm max}/\beta_{\rm max}^{\circ}$ and therefore $\xi$ increase
with $\Omega_{\rm psf}$. When reaching the spike regime (see
Sect.~\ref{lc}), $\beta_{\rm max}\rightarrow\beta^{\circ}_{\rm max}$
and $\xi\approx 1$. In this case, the decrease of the detectable event
cross section per source star, i.e. the area of the
$\beta\leq\beta_{\rm max}$ disc, is exactly compensated by the
increase of the number of such stars within the field of view.
\par In practice, however, the gradual increase of the detection rate
with $\Omega_{\rm psf}$ at constant number of resolution elements
$\Omega/\Omega_{\rm psf}$ is limited by the finite size of the stellar
sources and of the target galaxy. The finite source effect is
illustrated for example in Fig.~\ref{pixel} by the WHT/PFIP case,
where the growth of $N_{\rm det}$ with increasing surface brightness
reverses near the center because $\beta_{\rm max}$ approaches the
$\beta_{\rm min}$ limit. For observations by large telescopes in the
$K$-band, this limit is not reached and thus $N_{\rm det}$ does not
depend on our rough evaluation of $\beta_{\rm min}$. Note also that
$\omega_{\rm eff}^{-1}$ decreases with $\beta$, which means that an
improved time resolution is needed at degrading spatial resolution.
\par According to Fig.~\ref{pixel}, the number of detections in the
central region of Maffei~1 is superior for ground-based 8m-class
telescopes in the $K$-band than for HST in the $I$-band, but
$N_{\rm det}$ declines faster with radius in the $K$-band than in the
$I$-band. Table~\ref{pixres} gives the total number of $5\sigma$ event
detections when pointing the galaxy centre and integrating
$N_{\rm det}$ plotted in Fig.~\ref{pixel} over the field of view of
the detectors, which is the truly relevant quantity to inter-compare
the capabilities of the various observing means, as well as the
fraction of detections where the timescale can be inferred with a
precision better than 50\%. The predicted detection numbers are quite
substantial, with over 10~events for Gemini North and for HST if one
takes into account the factor 2 gained by interchanging the role of
the source and reference images. Even the other telescopes should all
yield of order one or more detections per double night. However, the
timescales may reasonably be accessed only by the largest telescopes
with AO in the $K$-band. In particular, Gemini North should provide
about one timescale at the 50\% precision level every two nights.
\par Note that according to the terminology and the $F_{\rm max}$
versus $F_*$ criteria developed in G96, pixel lensing of the central
region of Maffei~1 falls in the spike regime for WHT/PFIP, in the
semiclassical regime for HST/ACS, at the transition between the two
regimes for Keck/LRIS, and in the semiclassical regime for all
$K$-band observing facilities considered here.
\par The so far computed detection numbers are based on
$D_{\rm L}=10$~kpc, $v=50$~km\,s$^{-1}$ and
$m_{\rm L}=1\;M_{\oplus}$, implying $\omega^{-1}=13$~hours. The
weak dependence of the timescale on $D_{\rm L}$ in the context of
circular motion has already been highlighted in Sect.~\ref{times}, and
Eq.~(\ref{eqpix}) reveals that $N_{\rm det}$ depends on $\omega$ only
via the $\xi$ integral, which in the spike regime limit is also
independent of $\omega$. If $D_{\rm L}=3$~kpc while keeping $v$ and
$m_{\rm L}$ at the same default values,
the~$\int N_{\rm det}{\rm d}\Omega$ listed in Table~\ref{pixres}
change to 6.2 for HST/ACS and to 7.3 for Gemini/NIRI, showing that our
rather large default choice of $D_{\rm L}$ in fact underestimates the
true number of detections. If now $v=\{25,100\}$~km\,s$^{-1}$ while
keeping the other parameters at their default values, these numbers
become respectively $\{4.6,6.0\}$ for HST/ACS and $\{5.9,7.4\}$ for
Gemini/NIRI. These examples illustrate that introducing a distance and
a velocity distribution of the H$_2$ cores in our pixel lensing
detection calculation will not modify much the reported results.
\begin{table*}
\centering
\caption{\small
Characteristics of some telescope plus instrument
combinations, and the resulting expected total number of pixel lensing
events $\int N_{\rm det} {\rm d}\Omega$ over the full field of view
with a total signal-to-noise ratio $Q\geq 5$ and the fraction
$p(\delta w/w \leq 0.5)$ of these events with timescales measurable at
a precision better than $50\%$, when observing the central region of
Maffei~1 in a 10~hour program and assuming pre-existing reference
images at the same noise level as the event images. More precisely,
$p(\delta w/w\!\leq\!\varepsilon)\equiv
\int \Delta N_{\rm det}(\delta w/w\!\leq\!\varepsilon){\rm d}\Omega/
\int N_{\rm det}{\rm d}\Omega$.
AO stands for instruments in adaptive optic mode. $Z$ is the telescope
zeropoint magnitude yielding 1~e$^-\!$/s, and $S_{\rm sky}$ the
background comprising the sky and any instrumental background not
related to the observed galaxy. The adopted values refer to data found
on the instrument web pages (except for Keck and WHT $I$-band
backgrounds). The seeing $\theta_{\rm see}$ represents the full width
at half maximum of the point spread function. For the HST, the value
is taken from Buta \& McCall~\cite{BMC2}. With AO, the diffraction
limit $\theta_{\rm see}\approx 1.22 \lambda/D$, where $D$ is the
telescope diameter, is adopted, and otherwise the median site seeing
$\theta_{\rm see}\approx 0.7''(\lambda /[0.5 \mu{\rm m}])^{-1/5}$,
valid for Mauna Kea and La Palma, is assumed.}
\begin{tabular}{lccccccc} \hline\hline \vspace*{-.35cm} \\
Telescope/instrument & Filter & $Z$ & $S_{\rm sky}$ & $\theta_{\rm see}$ &
Field of view & $\int N_{\rm det}{\rm d}\Omega$ &
$p(\delta w/w \leq 0.5)$\\
& & (mag for 1 e$^-\!$/s) & [mag/arcsec$^2$] & [arcsec] & [arcsec$^2$] &
& [\%] \\
\vspace*{-.35cm} \\ \hline
HST/ACS         & $I$ & 25.5 & 21.43 & 0.076 & $202\times 202$& 5.11 & 3.9 \\
Keck/LRIS       & $I$ & 28.0 & 19.5  & 0.63 & $360\times 468$ & 1.92 & 2.5 \\
WHT/PFIP        & $I$ & 25.6 & 19.5  & 0.64 & $972\times 972$ & 0.46 &   0 \\
Gemini/NIRI(AO) & $K$ & 26.1 & 13.75 & 0.070 & $51\times 51$  & 6.32 & 8.7 \\
Subaru/CISCO    & $K$ & 26.1 & 12.7  & 0.52 & $108\times 108$ & 1.54 & 6.1 \\
CFHT/KIR(AO)    & $K$ & 24.0 & 11.6  & 0.154 & $36\times 36$  & 0.53 & 5.6 \\
WHT/INGRID      & $K$ & 24.5 & 12.0  & 0.52 & $244\times 244$ & 0.76 & 3.9 \\
\hline\vspace*{-0.01cm}
\label{pixres}
\end{tabular}
\end{table*}
\par The effect of varying $m_{\rm L}$ is two-folded. Firstly,
$\omega^{-1}\propto \sqrt{m_{\rm L}D_{\rm L}}/v$ and therefore this
parameter acts the same way as $D_{\rm L}$, producing a modest
increase of $\int N_{\rm det}{\rm d}\Omega$ at decreasing $m_{\rm L}$.
Secondly,
$\beta_{\rm min}\propto r_{\rm E}^{-1}\!\propto m_{\rm L}^{-1/2}$,
thus $\beta_{\rm min}$ rises with decreasing $m_{\rm L}$ and as a
consequence $N_{\rm det}$ diminishes. This decrease becomes important
especially when $\beta_{\rm min}$ gets close to $\beta_{\rm max}$,
which first happens at the galaxy centre. The net effect at decreasing
$m_{\rm L}$ is that $\int N_{\rm det}{\rm d}\Omega$ passes through a
maximum of 5.4 events at $m_{\rm L}\approx 0.2\;M_{\oplus}$ for
HST/ACS, and of 11.2 events at
$m_{\rm L}\approx 0.006\;M_{\oplus}$ for Gemini/NIRI, and
$N_{\rm det}$ starts to vanish at the centre at
$m_{\rm L}\approx 0.02\;M_{\oplus}$ for HST and
$0.0008\;M_{\oplus}$ for Gemini. The number of detections does
not depend much on the mass spectrum as long as most lenses have
masses above these critical values. However, at
$m_{\rm L}=0.001\;M_{\oplus}$, the timescale is only $\sim\! 50$~min,
requiring high frequency sampling.
\par If one includes those events with $\beta_{\rm max}>1$, i.e.
releases the lower limit of 1 in Eq.~(\ref{db}), then
$\int N_{\rm det}{\rm d}\Omega=5.2$ for HST/ACS and $8.3$ for
Gemini/NIRI. The gain is larger in the second case because the events
have larger $\beta$ at fixed value of $Q$. If now the detection
threshold is set to $Q_{\rm min}=7$, then
$\int N_{\rm det}{\rm d}\Omega$ would be 3.3 for HST/ACS and 4.9 for
Gemini/NIRI.
\par As the timescale decreases, the fraction of event with
$\delta w/w\leq 0.5$ increases because a larger portion of the light
curve is probed within the fixed observing time window $\Delta t$.
To illustrate this, taking $v=100$~km\,s$^{-1}$ reduces the timescale
by a factor of two and transforms the values of
$p(\delta w/w\leq 0.5)$ given in Table~\ref{pixres} to 9.3\% for
HST/ACS and to 18.6\% for Gemini/NIRI.
\par It should be noted that the $51''\times 51''$ field of view
assumed for NIRI/AO relies on the non-conventional but practically
possible f/14 camera mode. In the normal f/32 mode, the field of view
would be only $22''\times 22''$. Although optimal PSF sampling is not
a necessity for pixel lensing, at f/14, $\Omega_{\rm psf}$ is still
sampled by 6 pixels in the $K$-band. In all other selected telescope
and instrument combinations, the sampling is at least as good.
Figure~\ref{field} also displays some possible AO guide stars. The
central grey dot is in fact a 12.1 magnitude compact nuclear source
with an intrinsic full width at half maximum of $0.080''$ (Buta \&
McCall~\cite{BMC2}), and represents an acceptable option. Otherwise,
the best point source appears to be the 12.3 magnitude star located
$25''$ south-west from the galaxy centre. One pitfall of the current
Gemini North AO system, which explains the avoidance of the NIRI f/14
mode, is a degradation of the image quality beyond $\sim\! 10''$ from
the guide star. If this is a problem, one may still resort to the IRCS
instrument on the Subaru telescope, with a $58''\times 58''$ field of
view and where the AO ensures a good image quality within $30''$ from
the guide star.
\par Finally, regarding the noise induced by systematic effects
(time-variable PSF, photometric and geometric misalignments and
discrete pixelisation), G96 has shown that it can always be reduced
below the photon noise. Clearly, observations from space are free of
airmass constraints, and in particular will be less affected by
variable PSF and image distortion problems.

\section{Constraints from microlensing experiments}
\label{constr}

The EROS and MACHO collaborations (Alcock et al. \cite{EM}) have
published combined limits on the amount of planetary-mass dark matter
objects in the Galactic halo from their classical microlensing surveys
of the Magellanic Clouds, relying on halo models more spherical than
an ellipticity of E6. For halos composed entirely of Earth-mass
objects, they predict up to 100 lensing events whereas none was found,
and more generally, they conclude that objects with
$3.5\cdot 10^{-7}<m_{\rm L}/M_{\odot}<4.5\cdot 10^{-5}$ make up less
than 10\% of the total dark halo mass.
\par This does not rule out the possible existence of massive
H$_2$-core discs in the Milky Way and the LMC. Indeed, in the case of
LMC searches, $D_{\rm S}\approx 55$~kpc and typical lens distances and
transverse velocities are $D_{\rm L}=15$~kpc and
$v\sim 175$~km\,s$^{-1}$ for halo lenses
(Renault et al.~\cite{RAB}), $D_{\rm L}\!\sim\! 0.5$~kpc and
$v\!\sim\! 50$~km\,s$^{-1}$ for Galactic H$_2$ lenses, and
$(D_{\rm S}\!-\!D_{\rm L})\!\sim\! 0.3$~kpc and
$v\!\sim\! 30$~km\,s$^{-1}$ for LMC H$_2$ lenses (Gyuk et
al.~\cite{GDG}), so that at given lens mass,
$\omega^{-1}\propto[D_{\rm L}(D_{\rm S}-D_{\rm L})]^{1/2}/v$ does not
vary much more than \hbox{$\sim\!50\%$} among these different lens
phase-space distributions. Hence the classical detection rates, as
given by Eq.~(\ref{Go}), depend predominantly on the optical depths.
\par For a traditional roundish Galactic dark halo full of compact
lenses, $\tau\approx 5\cdot 10^{-7}$. For Galactic H$_2$ cores with
$\gamma=2$ times the mass density of the HI distribution, assuming
that the HI has a Gaussian vertical distribution with standard
deviation $\sigma$ and local surface density $\Sigma$ and that the Sun
sits in the middle of the Galactic plane, one gets:
\begin{equation}
\tau=\frac{2\sqrt{2\pi}G\gamma\Sigma\sigma}{c^2\sin^2{b}}.
\end{equation}
The Galactic latitude of the LMC is $b\approx -33^{\circ}$, and taking
$\sigma=300$~pc and $\Sigma=5\;M_{\odot}/{\rm pc}^2$, the result is
$\tau=2.4\cdot 10^{-9}$. For H$_2$ cores in the LMC disc, the optical
depth must be averaged over the stellar sources in this disc. Assuming
the same $\gamma$ factor as for the Milky Way, a Gaussian vertical
distribution of the LMC HI gas with $\sigma=180$~pc and an average
$\Sigma/\cos{i}=22.4\;M_{\odot}/{\rm pc}^2$ (Kim et al.~\cite{KDS}),
where $i\approx 35^{\circ}$ is the inclination angle of the LMC disc
from face-on, and an exponential vertical stellar distribution with a
scale height of 300~pc (Gyuk et al.~\cite{GDG}), one finds
\hbox{$\tau=5.1\cdot 10^{-9}$}. Hence the microlensing optical depths
of H$_2$ cores in the Milky Way disc and in the LMC disc are in both
cases two orders of magnitude less than in the case of lenses
distributed in a more spherical halo, rendering the expected detection
number of such cores consistent with the reported non-detection of
Earth-mass lenses.
\par A similar argument also holds for pixel lensing experiments
targeting M31. Indeed, the predicted optical depth towards this galaxy
due to $\sim\!$~spherical dark halos of compact objects around both
the Milky Way and Andromeda is $\tau\approx (5-10)\cdot 10^{-6}$
(e.g. Crotts~\cite{C}) and predicted lensing rates for
$10^{-3}\la m_{\rm L}/M_{\odot}\la 1$ reach about 200 event per year
(Kerins et al.~\cite{KCE}; Han~\cite{HC}), while the optical depth of
Galactic disc H$_2$ cores towards M31 is similar to that inferred
towards the LMC. According to Sect.~\ref{pl},
$\Gamma\sim \tau\cdot\xi$. When passing from
$(0.01\!-\!1)\;M_{\odot}$ halo lenses to Earth-mass Galactic
H$_2$-core lenses, $\tau$ decreases by over 3 orders of magnitude
whereas $\xi$ increases by no more than a factor of $\sim\! 20$,
because $\xi\sim \Delta\beta/\beta_{\rm max}^{\circ}\leq\zeta$ and
$\zeta$ increases from $\sim\! 0.05$ to 1 when $\beta$ goes from 1 to
0 if $t_1\!\rightarrow\! -\infty$ and $t_2\!\rightarrow\! +\infty$.
The last inequality particularly reflects the non-zero
$\beta_{\rm min}$ value. Hence the detection rate
of H$_2$ cores falls by a factor $\ga\! 100$, as for the LMC
experiments.
\par The MEGA microlensing team (Alves et al.~\cite{ABC}) has
attempted to put limits on low-mass halo objects from a high
time-resolution survey of M31 using the Subaru telescope, which
unfortunatley encountered bad weather. For halos with
100\% earth-mass objects, they predict about 60 events in two nights.
If the lenses were \hbox{H$_2$-cores} in the Galactic disc, this rate
would reduce to roughly 0.1 event per night, well below the
sensitivity of equivalent searches towards Maffei~1.
\par An important point is that by symmetry the absolute Einstein
radius is similar for lenses in the Milky Way and lenses in the source
galaxy. Referring to Eq.~(\ref{Bmin}), this means in particular that 
$\beta_{\rm min}\propto D_{\rm L}$ and therefore the low-mass lenses
in a distant source galaxy will have a negligible detection rate
relative to the local ones.

\section{Conclusion}
\label{concl}

The theory of cold H$_2$ globules with condensed central cores to
account for dark matter in spiral galaxies can be readily tested by
microlensing. If such globules exist in the Milky Way, the former
microlensing experiment have looked at directions which do not
optimise their detection rates. These globules are indeed expected to
follow the HI distribution and should be concentrated in the outer
Galactic disc, where the contribution of invisible mass is highest.
The large galaxies in the Maffei group provide excellent microlensing
targets to probe the H$_2$-core content of this region. In particular,
the Maffei~1 elliptical is at a distance of only 3~Mpc, very close to
the Galactic plane and $\sim\! 44^{\circ}$ away from the Galactic
anti-centre, and offers a huge reservoir of microlensing sources.
\par The condensed H$_2$-globule cores have mass and size comparable
to the Earth or less. The Earth-mass Galactic H$_2$ core--Maffei~1
lensing geometry implies an Einstein radius of $\sim\! 3\;R_{\odot}$,
large enough to consider the source stars as point-like over a large
range of impact parameters. Assuming that the H$_2$ cores are
distributed with twice the HI mass density, the microlensing optical
depth towards Maffei~1 of these cores is
$\tau\approx 0.7\cdot 10^{-6}$, and in the circular orbit
approximation, the typical timescale of the lensing events is at most
$\sim\! 1$~day.
\par In the classical lensing approach, due to the fluctuation noise,
the surface density of resolvable stars in Maffei~1 as a function of
apparent major-axis radius $a$ peaks away from the centre,
at $a\approx 230$~arcsec in the $I$-band and at $a\approx 90$~arcsec
in the $K$-band, and high resolution is crucial to maximise the
amplitude of this peak. Because of the smaller Maffei~1 area with high
density of resolvable stars and the restricted detector size in the
$K$-band, and of the poor seeing in ground-based optical observations,
the ACS camera on HST is currently the only useful instrument. In the
$I$-band and monitoring several fields at the same time, it would take
$\sim\! 5$ days of ACS observations to detect one event.
\par The pixel lensing approach, however, reveals over one order of
magnitude more efficient. In this case, the detection rate is maximum
in the central region ($a=0$) and it is best to rely on a single field
centred at this position. Since pixel lensing involves the subtraction
of reference images to the images with the ongoing lensing events, at
least two observing runs are required, spaced by a time interval large
relative to the lensing timescale, to get the two sets of images. The
roles of the reference and event images can fortunately be
interchanged to double the number of detectable events. We find that
in the $K$-band and resorting to adaptive optics, the ground-based
8m-class telescopes could detect as much as 10 events at the $5\sigma$
level in two nights, and that in the $I$-band, the ACS camera could
detect a similar number of events at the same threshold in two
10-hours observing periods.
\par About $9\%$ of the events detected by 8m-class telescopes in $K$
should have timescales measurable at a precision better than $50\%$,
representing about one such timescale per couple of nights, while this
fraction reduces to $\sim\! 4\%$ for the ACS observations in $I$. A
potentially powerful way to constrain the typical lens mass
$m_{\rm L}$ could be to measure the event detection rate as a function
of the radius~$a$. Indeed, the average impact parameter of detectable
events decreases with increasing galaxy surface brightness, and the
finite radius of the source stars imposes a lower limit
$\beta_{\rm min}\propto m_{\rm L}^{-1/2}$ below which the true flux
amplification becomes much less than the point-source approximation.
As a consequence, the detection rate curve will display a maximum at a
radius related to the lens mass, with precise location depending on
the signal-to-noise threshold.
\par The advent of instruments like WEBCam on JWST in the near-IR,
with a field of view of $2.16'\times 2.16'$, will boost the detection
numbers derived here. Although a positive detection will not yet prove
that the intervening lens is an H$_2$ core, it will certainly
represent a significant step in understanding the origin of dark
matter, whereas a no-detection result may refute the cored H$_2$
globules as a representative mass constituent of the Galaxy.
\begin{acknowledgements}
The author is thankful to Andr\'e Blecha for his informations
regarding observational aspects with large telescopes. The project
described in this paper will not be conducted by the author due to
lack of Swiss funding, but the author would appreciate to be
associated as co-investigator in its potential realisation.
\end{acknowledgements}

\end{document}